\documentclass[12pt]{article}
\usepackage{amssymb}
\makeatletter
\if@titlepage

  \renewcommand\maketitle{\begin{titlepage}%
  \let\footnotesize\small
  \let\footnoterule\relax
  \null\vfil
  \vskip 60\p@
  \begin{center}%
    { \bfseries \LARGE \@title \par}%
    \vskip 3em%
    {\normalsize
     \lineskip .75em%
      \begin{tabular}[t]{c}%
        \@author
      \end{tabular}\par}%
      \vskip 1.5em%
    {\normalsize \@date \par}
  \end{center}\par
  \@thanks
  \vfil\null
  \end{titlepage}%
  \setcounter{footnote}{0}%
  \let\thanks\relax\let\maketitle\relax
  \gdef\@thanks{}\gdef\@author{}\gdef\@title{}}
\else
\renewcommand\maketitle{\par
  \begingroup
    \renewcommand\thefootnote{\fnsymbol{footnote}}%
    \def\@makefnmark{\hbox to\z@{$\m@th^{\@thefnmark}$\hss}}%
    \long\def\@makefntext##1{\parindent 1em\noindent
            \hbox to1.8em{\hss$\m@th^{\@thefnmark}$}##1}%
    \if@twocolumn
      \ifnum \col@number=\@ne
        \@maketitle
      \else
        \twocolumn[\@maketitle]%
      \fi
    \else
      \newpage
      \global\@topnum\z@   
      \@maketitle
    \fi
    \thispagestyle{plain}\@thanks
  \endgroup
  \setcounter{footnote}{0}%
  \let\thanks\relax
  \let\maketitle\relax\let\@maketitle\relax
  \gdef\@thanks{}\gdef\@author{}\gdef\@title{}}
\def\@maketitle{%
  \newpage
  \null
  \vskip 1em%
  \begin{center}%
    {\Large \bfseries\@title \par}%
    \vskip 1.5em%
    {\small
      \lineskip .5em%
      \begin{tabular}[t]{c}%
        \@author
      \end{tabular}\par}%
    \vskip 1em%
    {\normalsize \@date}%
  \end{center}%
  \par
  \vskip 1.5em}
\fi
\renewcommand\section{\@startsection{section}{1}{\z@}%
                                     {-3.25ex\@plus -1ex \@minus -.2ex}%
                                     {1.5ex \@plus .2ex}%
                                     {\reset@font\normalsize\bfseries}}
\renewcommand\subsection{\@startsection{subsection}{2}{\z@}%
                                    {3.25ex \@plus1ex \@minus.2ex}%
                                    {-1em}%
                                    {\reset@font\normalsize\bfseries}}
\@addtoreset{equation}{section}
\makeatother
  \def\R{\mathbb{R}}

\newcommand{\rnote}[1]{\raisebox{1ex}{{\hspace*{-3mm} \scriptsize\sf#1}}
                       \hspace*{-4mm}}

\def\be{ \begin{equation}}          \def\ee{ \end{equation}}
\def\ba{ \begin{eqnarray}}          \def\ea{ \end{eqnarray}}
\def\nn{\nonumber}                  
\def\ve{\varepsilon}

\def\Im{{\rm Im\,}}

               \def\ve{\varepsilon}
           \def\s{\sigma}
\def\la{\lambda}

\title{Moduli Spaces \\[2mm] of D-branes in CFT-backgrounds}
\author{{\sc Andreas Recknagel \rnote{1}\ , 
           \ Volker Schomerus \rnote{2,} \ \ \rnote{3} } \\[5mm] 
\rnote{1}\ \   
MPI f\"ur Gravitationsphysik, Albert-Einstein-Institut\\
Schlaatzweg 1, D-14473 Potsdam, Germany\\[4mm] 
\rnote{2}\  \ II. Inst. f. Theor. Physik, Universit\"at Hamburg\\
Luruper Chaussee 149, D-22670 Hamburg, Germany \\[4mm]
\rnote{3}\ \ Institut Mittag-Leffler, Aurav\"agen 17\\
S-182 62 Djursholm, Sweden} 
\date{March 3, 1999}
\begin{document}
\begin{titlepage}      \maketitle       \thispagestyle{empty}

\vskip1cm
\begin{abstract} \noindent 
D-branes in curved backgrounds can be 
treated with techniques of boundary conformal field 
theory. We discuss the influence of scalar condensates
on such branes, i.e.\ perturbations of boundary conditions
by marginal boundary operators. A general criterion is
presented that guarantees a boundary perturbation to be
truly marginal in all orders of perturbation theory. 
Our results on boundary deformations have several
interesting applications which are sketched at the
end of this note. 
\end{abstract}
\vspace*{-17.8cm}
{\tt {REPORT No.\ 12, 1998/1999 \hfill hep-th/9903139\ \ \ }}
\\
{\tt {DESY 99-013\hfill \phantom{hep-th/9903139}}}
\vfill
\noindent
\phantom{wwwx}\hspace*{1cm} {\small e-mail: }
 {\small\tt  {vschomer@x4u.desy.de, 
\  anderl@mis.mpg.de}}
\vskip.5cm
\noindent{\small Talk given by V. Schomerus at the 
$32^{\rm nd}$ International Symposium Ahrenshoop\\ on the Theory of 
Elementary Particles (Buckow, Germany, September 1998)}  
\end{titlepage}
\phantom{a} 
\setcounter{page}{0} 
\thispagestyle{empty} 
\newpage  

\phantom{a} \vspace*{1cm} 
It was observed many years ago that low energy effective 
actions of $\mbox{(super-)}$ string theories possess solitonic solutions.
They are known as solitonic p-branes because 
of their localization along certain $p+1$-dimensional surfaces 
in the string-background. Later, Polchinski discovered a
remarkable correspondence between such solitonic $p$-branes
and D$p$-branes, i.e.\ boundary conditions for open strings
(for a review see \cite{Pol}). This makes it possible to study 
branes through the `microscopic' techniques of boundary conformal 
field theory. The latter are particularly effective in dealing 
with curved backgrounds and they have also been successfully 
applied to study non-BPS branes. 
\medskip%

We consider conformal field theories on the upper half of the complex 
plane. Any such theory contains two different kinds of fields, namely 
{\em bulk fields} $\phi(z, \bar z)$ (i.e.\ `closed string vertex 
operators') which are defined for points in the interior $\Im z >0 $ 
of the upper half-plane and {\em boundary fields} $\psi(x)$ (i.e.\ 
`open string vertex operators') that are inserted along the boundary 
$\Im z=0$. Operator product expansions (OPEs) of the former specify 
the background of our string theory. Even if we do not allow for 
changes of these {\sl bulk OPEs} there remains some freedom to 
choose different {\sl boundary conditions}. Roughly speaking, this 
corresponds to the freedom of placing D-branes in a fixed background. 
A somewhat axiomatic characterization of the admissible boundary 
theories can be found e.g.\ in \cite{ReSc}.  
\medskip

The problem of constructing boundary theories possesses some 
universal solutions in terms of the modular $S$-matrix, which 
were found by Cardy in \cite{Car}. They provide at least a finite 
set of boundary conformal field theories (D-branes) for a 
large class of backgrounds. On the other hand the relation 
with solitonic p-branes suggests the existence of continuous 
families of solutions. We shall explain below how such 
families may be recovered from a discrete set of boundary 
CFTs. The idea is to generate continuous deformations through 
perturbations with marginal boundary fields. Since the 
boundary operator content depends on the boundary condition, 
the number of possible deformations will depend on the boundary 
theory we start with.
\medskip

Suppose we are given a set of correlation functions of some 
boundary conformal field theory. We want to perturb this 
initial theory by one of its boundary fields $\psi(x)$ 
according to the formal prescription
\ba
\lefteqn{\langle\, \varphi_1(z_1,\bar z_1) \cdots 
    \varphi_N(z_N, \bar z_N)\,\rangle_{ \lambda\psi} 
  =   Z^{-1} \cdot \langle\, I_{\lambda\psi} \, \varphi_1(z_1,\bar z_1) 
 \cdots \varphi_N(z_N, \bar z_N) \,\rangle }
\label{defcorra} \nn \\[2mm] 
& := & Z^{-1} \sum_n \frac{\la^n}{n!}  \sum_{\s \in S_n}
       \int \cdots \int_{\raisebox{-3mm}{$ \scriptstyle 
        x_{\s(i)}\, <\, x_{\s(i+1)}$}} 
       \hspace*{-15mm} \frac{dx_1}{2\pi} \cdots \frac{dx_n}{2\pi} 
       \langle\, \psi(x_{\s(1)}) \cdots \psi(x_{\s(n)} ) 
     \  \varphi_1 \cdots 
    \varphi_N \,\rangle  
      \nn   \ea
where $\lambda$ is a real parameter.The symbol $I_{\lambda \psi}$ 
in the first line should be understood as a path ordered exponential
of the perturbing operator,  
\be  I_{\lambda\psi} \ = \ P \exp \bigl\lbrace\, \lambda 
   S_\psi\,\bigr\rbrace \ :=\  P \exp \bigl\lbrace\, 
    \lambda \int_{- \infty}^\infty \psi(x) \, 
     \frac {dx} {2\pi}\,\bigr\rbrace\ \  .  
\label{defcorrb}\ee
The normalization $Z$ is defined as the expectation value 
$Z \sim \langle I_{\la \psi}\rangle$. These expressions 
deal with deformations of bulk correlators only. If there are extra 
boundary fields present in the correlation function, the formulas 
need to be modified in an obvious fashion so that these boundary 
fields are included in the ``path ordering''. Some particularly 
simple cases of this type will be discussed shortly. 
The path ordering is essential since correlations of boundary 
fields are only defined for an ordered set of insertion points 
along the real line. Their analytic continuation beyond this domain 
is not unique, in general.  

To make sense of the above expressions (beyond the formal level), it is 
necessary to regularize the integrals (introducing UV- and 
IR-cutoffs) and to renormalize couplings and fields (for an 
introduction into bulk perturbations in 2D conformal field 
theories see e.g. \cite{CarL}). IR divergences 
can be cured by putting the system into a ``finite box''. On the other 
hand, we have to deal with UV divergences. So, let us introduce a UV 
cutoff $\ve$ such that the integrations are restricted to the region 
$|x_i - x_j|>\ve$. Thereby, all integrals become UV-finite before we 
perform the limit $\ve \rightarrow 0$.%
\medskip%

In the following, we consider marginal boundary deformations 
where the conformal dimension $h$ of the perturbing operator
$\psi(x)$ is $h=1$ so that there is a chance to stay at the 
conformal point for arbitrary values of $\lambda$. Of course,
the dimension of the field $\psi$ may decrease upon
perturbation to values $h < 1$. In such cases, the
perturbation with $\psi(x)$ is said to be {\em marginally
relevant} and the perturbed theories are no longer
scale invariant. Here we are interested in a condition
which guarantees that $\psi(x)$ is {\em truly marginal},
i.e.\ that the dimension of $\psi(x)$ stays at its
initial value $ h = 1 $. Perturbations with truly
marginal fields then lead to a one-parameter family 
of scale invariant theories.
\smallskip 

It was shown in \cite{ReSc} that {\em $\psi(x)$ is truly
marginal provided that it is a self-local field
of conformal dimension $h = 1$}. The condition of 
self-locality requires $\psi$ to obey
$$ \psi(x)\  \psi(y) \ = \ \psi(y)\  \psi(x) \ \
\mbox{ for } \ \ x < y\ .
$$
Here we assumed that the right hand side can be
unambiguously defined by analytic continuation.
\smallskip

Let us briefly sketch the main idea behind this result. Because 
of its self-locality, $\psi(x)$ behaves like a connection in an 
abelian gauge theory so that we may employ the usual arguments 
to omit the path-ordering in our construction of the deformed 
correlators. Consequently, all  
integrals can be performed along the whole real axis. The 
cutoff $\ve$ together with the normalization by $Z$ results 
in a small shift of the integration contour into the upper 
half-plane. Since these contour integrals do no longer depend 
on the cutoff, the limit $\ve \rightarrow$ becomes trivial, 
i.e.\  
\ba
\lefteqn{
\langle\, \varphi_1(z_1,\bar z_1) \cdots 
    \varphi_N(z_N, \bar z_N)\,\rangle_{ \lambda\psi}
= \lim_{\ve\to 0} \; \langle\, \varphi_1(z_1,\bar z_1) \cdots 
    \varphi_N(z_N, \bar z_N)\,\rangle^{\ve}_{\lambda\psi}}
\label{analdef} \\[2mm]
& \phantom{xxxxx} = & \sum_n \frac{\la^n}{n!}  
       \int_{\gamma_1} \cdots \int_{\gamma_n}
        \frac{dx_1}{2\pi} \cdots \frac{dx_n}{2\pi} 
       \langle\, \psi(x_1) \cdots \psi(x_n ) 
     \  \varphi_1 \cdots 
    \varphi_N \,\rangle 
      \nn   \ea  
where $\gamma_p$ is the straight line parallel to the real 
axis with $\Im \gamma_p = i \ve/p $. The expression on the 
right hand side is manifestly finite and is independent 
of $\ve$ as long as $\ve <$ {\rm min\/}$(\Im z_i)$ where $z_i$ 
denote the insertion points of bulk fields. Thus, the above 
formula allows us to construct the perturbed bulk correlators 
to all orders in perturbation theory.
\smallskip

The extension of these ideas to the deformation of
correlators which involve boundary fields $\psi_1, \dots,
\psi_N$ is straightforward {\em if and only if} the boundary 
fields $\psi_1,\ldots,\psi_M$ are {\em local} with respect to 
the perturbing field $\psi$, i.e.\ iff
$$ \psi(x)\ \psi_i(y) \ = \ \psi_i(y) \ \psi(x) \ \
\mbox{ for } \ \ x < y\ .
$$
This usually places a strong constraint on boundary fields.
But if it is satisfied, the (renormalized) correlation
functions are again obtained through contour integration
formulas very similar to the eqs.\ (\ref{analdef})
(see \cite{ReSc} for details). The explicit expressions
imply in particular that {\em the conformal dimension of a
boundary field $\psi'(x)$ is not changed upon deformation
provided that $\psi'$ is local with respect to $\psi$}.   
\medskip

The preceding statement has a number of nice consequences.  
First of all it certainly allows us to understand that non-locality
is the only obstruction to true marginality. Indeed, with 
the choice $\psi' = \psi$ our result implies that
the dimension of the perturbing self-local field $\psi$
sticks to its initial value $h =1$, i.e.\ that $\psi$ is
truly marginal.
\smallskip

Whenever we have several self-local boundary fields which are all
local with respect to each other, they generate a multi-parameter 
deformation of the initial boundary theory. At first sight, this
may seem to result in very trivial moduli spaces. 
However, it may happen that certain important properties
of the boundary conformal field theory (e.g.\ the cluster
property) break down at some finite value of the perturbation 
parameter $\lambda$ \cite{ReSc}: Admissible deformations
form a subspace $\Omega \subset {\R}^q$. The moduli 
space ${\cal M}$ we assign to our initial boundary condition  
is obtained from $\Omega$ upon identification of equivalent 
boundary theories, i.e.\ ${\cal M} = \Omega/ \sim$. Here, 
equivalence must be tested by closed string exchange between 
different branes, as explained in \cite{ReSc}. 

Moduli spaces of $D0$-branes are expected to probe the background
geometry, i.e.\ they provide a way of recovering the background 
from the purely algebraic CFT description. A number of explicit 
examples are discussed in \cite{ReSc}. The resulting geometry, 
however, is not unique due to the freedom one has in declaring 
a given boundary condition a $D0$-brane. Such choices may be 
understood as a consequence of T-duality (or mirror symmetry). 

For boundary conditions in Gepner models \cite{Gep} (i.e.\ 
D-branes on certain complete intersection Calabi Yau spaces), 
the moduli spaces ${\cal M}$ should come equipped with 
additional  K\"ahler structures much as their cousins in 
algebraic geometry (see e.g.\ \cite{SYZ}) but this and
other  predictions of algebraic geometry remain to be 
investigated in more detail. 
\smallskip%

To state another corollary of our main result we consider 
N identical branes placed on top of each other. Within the 
CFT framework this means that we multiply all boundary 
fields with Chan-Paton matrices, as usual. Thus, instead of 
perturbing our theories with $\lambda \psi$, we generate 
perturbations by $\Lambda \psi$ where  $\Lambda$ is now 
some N $\times$ N matrix. A single self-local boundary 
field $\psi$ can be combined with several different matrix 
parameters $\Lambda_i$. This gives rise to a multi-parameter 
deformation provided that the matrices $\Lambda_i$ commute,   
\be [\,  \Lambda_i\,  ,\, \Lambda_j\, ]\  = \ 0 \ \ .
\label{commu} \ee
We conclude that any self-local boundary field can generate an 
N-parameter deformation. In the case of D-branes on ${\R}^D$, 
this fact was first observed by Witten \cite{Wit}. The N parameters 
are then interpreted as distances between the N branes in a 
fixed direction transverse to their world-volume. In the 
present context, equation (\ref{commu}) makes a rather general 
statement about flat directions of the brane's low enery 
effective action in curved backgrounds. 
\smallskip

Deformations can have rather drastic effects on the geometry -- and 
in particular on the `dimension' -- of branes \cite{ReSc}. For a 
compactified bosonic theory at the self-dual radius there exist
boundary deformations that continuously deform Neumann- into 
Dirichlet-type boundary conditions. This phenomenon extends 
to other radii by means of radius-changing bulk deformations. Circle 
theories at rational radii possess additional tachyon condensates which 
are even capable of transforming Neumann-type branes into a superposition 
of Dirichlet-type branes and vice versa \cite{ReSc} (see also the 
recent work of Sen \cite{Sen}).%
\medskip

Let us finally mention that our formula (\ref{analdef}),  
and its appropriate generalization in the presence of 
boundary fields, provide a powerful tool 
in computing the deformed theory explicitly. They might 
be quite useful in studying certain bound states of D-branes. 
Related calculations by Sen (see \cite{Sen} and references 
therein) were argued in \cite{WitK} to support the idea of 
classifying D-branes through an appropriate K-theory of 
the background.


\begin{thebibliography}{77}
\bibitem{Pol} J.\ Polchinski, {\em TASI-lectures
 on D-branes}, hep-th/9611050  
\bibitem{ReSc} A.\ Recknagel, V.\ Schomerus, 
 {\it Boundary deformations and D-brane moduli},
 hep- th/9811237, Nucl. Phys. {\bf B} , to appear
\bibitem{Car} J.\ Cardy, {\em Boundary conditions,
 fusion rules and the Verlinde formula}, Nucl.\
 Phys.\ {\bf B 324} (1989) 581
\bibitem{CarL} J.\ Cardy, {\sl Conformal invariance and
 statistical mechanics},  Lectures given at the Les
 Houches Summer School in Theoretical Physics, 1988   
\bibitem{SYZ} A.\ Strominger, S.-T.\ Yau, E.\
 Zaslow, {\em Mirror symmetry is T-duality},
 Nucl.\ Phys.\ {\bf B 479} (1996) 243,
 hep-th/9606040  
\bibitem{Gep} A.\ Recknagel, V.\ Schomerus,
 {\sl D-branes in Gepner models}, Nucl.\ Phys.\
 {\bf B 531} (1998) 185, hep-th/9712186 \\[2mm] 
 M.\ Gutperle, Y.\ Satoh, {\em D-branes in
 Gepner models and supersymmetry}, hep-th/9808080;   
 {\em D0-branes in Gepner 
 models and N=2 black holes}, hep-th/9902120 
\bibitem{Wit} E.\ Witten, {\it Bound states of 
 strings and D-branes}, Nucl.\ Phys.\ {\bf B 460}
 (1996) 335
\bibitem{Sen} A.\ Sen, {\sl SO(32) spinors of type 
 I and other solitons on brane - anti-brane pair},
 JHEP {\bf 09} (1998) 023, hep-th/9808141;  {\em 
 Descent relations among bosonic D-branes}, 
 hep-th/9902105 
\bibitem{WitK} E.\ Witten, {\it D-branes and
 K-theory}, hep-th/9810188  
\end{thebibliography}
\end{document}